\documentclass[prd, showpacs]{revtex4}

\usepackage{amsfonts}
\usepackage{amsmath,amssymb}
\usepackage[small,bf]{caption}
\usepackage{graphicx}

\newcommand{\lb}[0] { \left( }
\newcommand{\rb}[0] { \right) }
\newcommand{\beqs} { \begin{eqnarray} }
\newcommand{\eeqs} { \end{eqnarray} }
\newcommand{\bsub} { \begin{subequations} }
\newcommand{\esub} { \end{subequations} }
\newcommand{\nn} {\nonumber}
\newcommand{\ep}[0] { \epsilon }

\newcommand{\EE}[2] {#1 \times 10^{#2}}
\newcommand{\degree}{\ensuremath{^\circ}}

\begin{document}

\title{On the relation between the neutrino flux from Centaurus A and the associated diffuse 
neutrino flux}

\author{Hylke B. J. Koers}
\email{hkoers@ulb.ac.be}
\author{Peter Tinyakov}
\affiliation{Service de Physique Th\'eorique, Universit\'e Libre de Bruxelles (U.L.B.),
CP225, Bld. du Triomphe, B-1050 Bruxelles, Belgium}

\begin{abstract}
Based on recent results obtained by the  Pierre Auger Observatory (PAO), it has been
hypothesized that Centaurus A (Cen A) is a source of ultra-high-energy
cosmic rays (UHECRs) and associated neutrinos.
We point out that the diffuse neutrino flux may be used to constrain the
source model if one assumes that the ratio between the UHECR
and neutrino fluxes outputted by Cen A is representative for other
sources.
Under this assumption we
investigate the relation between the neutrino flux
from Cen A and the diffuse neutrino flux.
Assuming furthermore that Cen A is the source of two UHECR events observed
by PAO,
we estimate the all-sky diffuse neutrino flux to be
$\sim$$200-5000 $ times larger than the
neutrino flux from Cen A. 
As a result, the diffuse neutrino fluxes associated with some of the
recently proposed models of UHECR-related neutrino production in Cen A
are above existing limits. Regardless of the underlying source model, our results indicate
that the detection of neutrinos from Cen A
without the accompanying diffuse flux would mean
that Cen A is an exceptionally efficient neutrino source.
\end{abstract}


\preprint{ULB-TH/08-03}

\maketitle

\section{Introduction}
The Pierre Auger Observatory (PAO) has recently reported 
new results \cite{Cronin:2007zz, Abraham:2007si}
on the arrival directions of the highest-energy cosmic rays (CRs).
The data show strong
evidence for anisotropy of these CRs, which suggests that at least
some sources of ultra-high-energy cosmic rays (UHECRs) are relatively
close. 
 The cosmic-ray anisotropy has been confirmed by other studies
using different statistical methods and source catalogs
\cite{Kashti:2008bw,2008MNRAS.388L..59G,Ghisellini:2008gb}.
The correlation between the arrival directions of these CRs
and the positions of known active galactic nuclei (AGNs) has lead the
 PAO to suggest that nearby AGNs, or astrophysical objects
with a similar spatial distribution, are the sources of
UHECRs \cite{Cronin:2007zz}.  However, the
observed deficit of UHECRs from the nearby Virgo cluster appears to be
incompatible with such a source distribution
\cite{Gorbunov:2007ja}. Therefore the origin of UHECRs remains unclear
at present.

Independent of a possible general connection between UHECRs and AGNs,
the PAO data raise the possibility that Centaurus A (Cen A) is
a source of UHECRs. 
The PAO analysis \cite{Cronin:2007zz, Abraham:2007si} associates
two events with Cen A, but it has been pointed out
that at least four events can be associated with Cen A if one takes account
of its morphology \cite{Moskalenko:2008iz}.
At a distance of $\sim$3.5 Mpc, Cen A (NGC 5128) is the nearest active
galaxy (see Ref. \cite{1998A&ARv...8..237I} for a review).
It is classified as a Fanaroff-Riley type I radio galaxy,
possibly harboring a misdirected BL Lac nucleus.
The galaxy is
believed to be powered by accretion on a $\sim$$10^8 M_{\odot}$
central black hole \cite{2005AJ....130..406S}.
It has a very compact nucleus, a pronounced northern jet, a
dimmer southern jet, and giant radio lobes extending out to $\sim$250
kpc.  
We likely observe the jets from the side, under a viewing angle of $50\degree - 80\degree$
\cite{2001AJ....122.1697T} (see, however, Ref. \cite{Hardcastle:2003ye}). Due to its proximity, it has been
suggested long ago that this galaxy may be the source of UHECR events
observed at Earth \cite{1978A&A....65..415C, Anchordoqui:2002hs,Romero:1995tn}.

A general prediction of models of UHECR acceleration is the
production of neutrinos through the interaction of 
accelerated protons (or nuclei) with the ambient photon field or with
target protons in the source. Hence, for a
given acceleration mechanism the neutrino and UHECR fluxes are
related. Detection of neutrinos from Cen A or limits on their
flux may therefore translate into constraints on the underlying
acceleration models.

Neutrino production in AGNs has been studied by many authors, see
e.g. Refs.
\cite{1981MNRAS.194....3B,1979ApJ...232..106E,Szabo:1994qx,Mannheim:1998wp,
Mannheim:1995mm,Silberberg:1979hd,Halzen:1997hw,Stecker:1991vm,Muecke:2002bi,Stecker:1995th,
1992A&A...260L...1M}. More recently, the authors of Refs.
\cite{Cuoco:2007qd, Kachelriess:2008qx} have presented estimates on
the neutrino flux from Cen A under the assumption that two out of the
27 UHECR events in the  PAO analysis can be attributed to this galaxy. These
authors have however not considered the diffuse flux due to all
(unresolved) neutrino sources within their models. Assuming that the
environment in Cen A is somehow representative for all sources of
UHECRs and accompanying neutrinos, the diffuse neutrino flux may also
constrain the source model. In this work we investigate the connection
between the UHECR-related neutrino flux from Cen A and the associated
diffuse neutrino flux.

In relating the diffuse flux to that from Cen A, we assume 
(as a working hypothesis) that Cen A is a `typical' source of
UHECRs and neutrinos, i.e. we assume universal 
UHECR and neutrino injection spectra with a fixed relative strength.
We do not make
any assumptions on the intrinsic luminosity or on the distance of the
UHECR sources.  Within the assumption of typicality, the diffuse
neutrino flux can be estimated by upscaling the neutrino flux from Cen
A using CR data.  The scaling factor is the product of a trivial factor standing for the
fraction of observed UHECR events that is attributed to Cen A, and a
non-trivial factor that accounts for the difference in CR and
neutrino mean free paths: as the UHECR flux from far-away sources is
strongly attenuated by interactions with the cosmic microwave
background, the ratio of the diffuse neutrino flux to the Cen-A
neutrino flux will be much larger than the observed ratio of the
diffuse UHECR flux to the Cen-A UHECR flux.  In this study we estimate
this scaling factor without relying on any specific source
model. For definiteness we assume that the UHECRs are protons,
the composition of UHECRs still being under debate
\cite{Arisaka:2007iz}.

We find that the all-sky diffuse neutrino flux is expected to be
$\sim$$200-5000 $ times larger than the
neutrino flux from Cen A, depending most strongly on the assumed source
evolution.
As a consequence,  diffuse neutrino fluxes associated with some of the
recently proposed models of UHECR-related neutrino production in Cen A
overshoot existing bounds by the AMANDA-II \cite{Achterberg:2007qp, Gerhardt:2007zz}
and  PAO \cite{Abraham:2007rj} experiments. Regardless of the underlying production model, our results indicate
that the detection of neutrinos from Cen A
without the accompanying diffuse flux would  imply
that Cen A is an exceptionally strong neutrino source.

This paper is organized as follows. In section 
\ref{sec:att} we discuss attenuation of the UHECR proton
and neutrino fluxes and estimate the ratio of the diffuse neutrino
flux to the  Cen-A neutrino flux.
In section \ref{sec:srcex} we apply these results to models
that were recently proposed in Refs. \cite{Cuoco:2007qd,Kachelriess:2008qx}. 
We summarize and discuss our work in section \ref{sec:conclusion}.
The appendices contain additional information on the computer
code used to calculate attenuation of the cosmic proton flux,
and an estimate of the neutrino effective area of the
IceCube experiment.

\section{Relating the diffuse proton and neutrino fluxes}
\label{sec:att}
The flux of high-energy protons from a cosmic source is attenuated
by redshift and by interactions with CMB photons. As a result, the
observed flux of UHECR protons is significantly smaller than the
flux that is injected by all sources. Neutrinos, on the other hand, only suffer redshift
energy losses, which is of far lesser importance. 
This difference boosts the diffuse neutrino flux reaching Earth compared
to the diffuse flux of UHECRs, an effect that should be taken into account
when normalizing the diffuse neutrino flux to the observed UHECR flux.
Under the assumption of a universal relation between the outputted neutrino and proton fluxes,
this effect may be parameterized by a parameter $H$ that we introduce in this section.

\subsection{The neutrino boost factor}

The observed differential flux $\phi$ from a single source at proper
distance $D$ (redshift $z$) can be expressed as
\beqs
\phi (E) = \frac{j^0 (E_0)}{4 \pi D^2 (1+z)} \frac{d E_0}{d E} \, ,
\eeqs
where $E$ is the observed energy, $E_0 = E_0 (E,z)$ is the
energy at the source, and $j^0$ denotes
the differential spectrum at
the source. Integrating over a cosmological distribution of sources,
the diffuse flux is equal to
(see, e.g., Ref. \cite{Berezinsky:2002nc})
\beqs
\label{eq:diffflux}
\phi^{\rm diff} (E) =   \frac{c n_0}{4 \pi} \int_0^{\infty} dz \,  \left| \frac{dt}{dz} \right| 
\frac{d E_0}{d E}  \ep(z) j^0 (E_0) \, ,
\eeqs
where $n_0$ is the present source density and $\ep(z)$ parameterizes
source evolution (no evolution corresponds to $\ep(z) \equiv 1$).
In this expression we assume that all sources are identical.
Within the
$\Lambda$CDM concordance model that we adopt,
\beqs
\left| \frac{dt}{dz} \right|  =
\frac{1}{ H_0 (1+z) \sqrt{\Omega_m (1+z)^3 + \Omega_{\Lambda}}} 
\, ,
\eeqs
where
$H_0 = 73$ km s$^{-1}$ Mpc$^{-1}$ denotes the present Hubble constant,
$\Omega_m = 0.24$ is the present matter density parameter, and 
$\Omega_{\Lambda} = 0.76$ is the present vacuum energy density parameter
\cite{Yao:2006px}. 

Since we assume a universal neutrino injection spectrum in this work,
we approximate the observed diffuse neutrino flux by
$\phi_\nu^{\rm diff} (E) \propto j_{\nu}^0(E)$.
(In the case of spectral breaks this neglects smearing due to
different source redshifts. We will come back to this issue in the next section.)
We normalize the diffuse neutrino flux to the integral UHECR flux $\Phi^{\rm diff}_p$
above a threshold energy $E_{\rm th}$, and define a constant of proportionality $H$ as follows:
\beqs
\label{eq:def:Eta}
\frac{\phi_\nu^{\rm diff} (E)}{j_{\nu}^0(E)} =  H (E_{\rm th}) \frac{  \Phi_p^{\rm diff} (E_{\rm th})}{J_p^0 (E_{\rm th})}
 \, ,
\eeqs
where $j^0_\nu$ is the differential neutrino  spectrum at the
source, and $J_p^0$ is the integral UHECR proton spectrum at
the source  (we will use capital symbols to refer to integral
spectra and fluxes, and lower-case symbols for differential ones).
The effect of the different UHECR proton and neutrino mean free
path lengths is now contained in the
scaling factor $H$, which we will refer to as the
neutrino boost factor. It can be determined for any
neutrino and any proton injection spectrum
from eq. \eqref{eq:diffflux}, together with a formula for $E_0 (E,z)$.
Note that $H$ depends on the threshold energy $E_{\rm th}$ but not on
the neutrino energy $E$.

We have computed the parameter $H$ numerically for power-law
proton and neutrino injection spectra,
$j^0_p (E) \propto E^{-p_p}$ and
$j^0_\nu (E) \propto E^{-p_\nu}$, respectively.
This is done with the help of a computer code that 
solves eq. \eqref{eq:diffflux} for protons and neutrinos and then determines $H$ from eq. \eqref{eq:def:Eta}.
Proton energy losses are taken into account in the continuous loss approximation
using expressions for the energy-loss time that are given in appendix \ref{app:beta0}.
In this process we integrate over a cosmological distribution of sources
up to redshift $z=5$. 
Because we do not know the redshift evolution of UHECR sources,
we consider as limiting cases both no evolution
and strong evolution tracing the AGN luminosity density evolution given in Ref. \cite{1998MNRAS.293L..49B}, i.e.
$\ep(z) \propto (1+z)^{3.4}$ up to $z=1.9$,
a constant $\ep$ up to $z=3$, and negative evolution 
$ \ep(z) \propto (z-3)^{-0.33}$ beyond
$z=3$.
We have not explicitly considered milder source evolution models, but the results
for $H$ will be numerically intermediate between the two cases considered.

\begin{figure}
\includegraphics[width=5cm, angle=270]{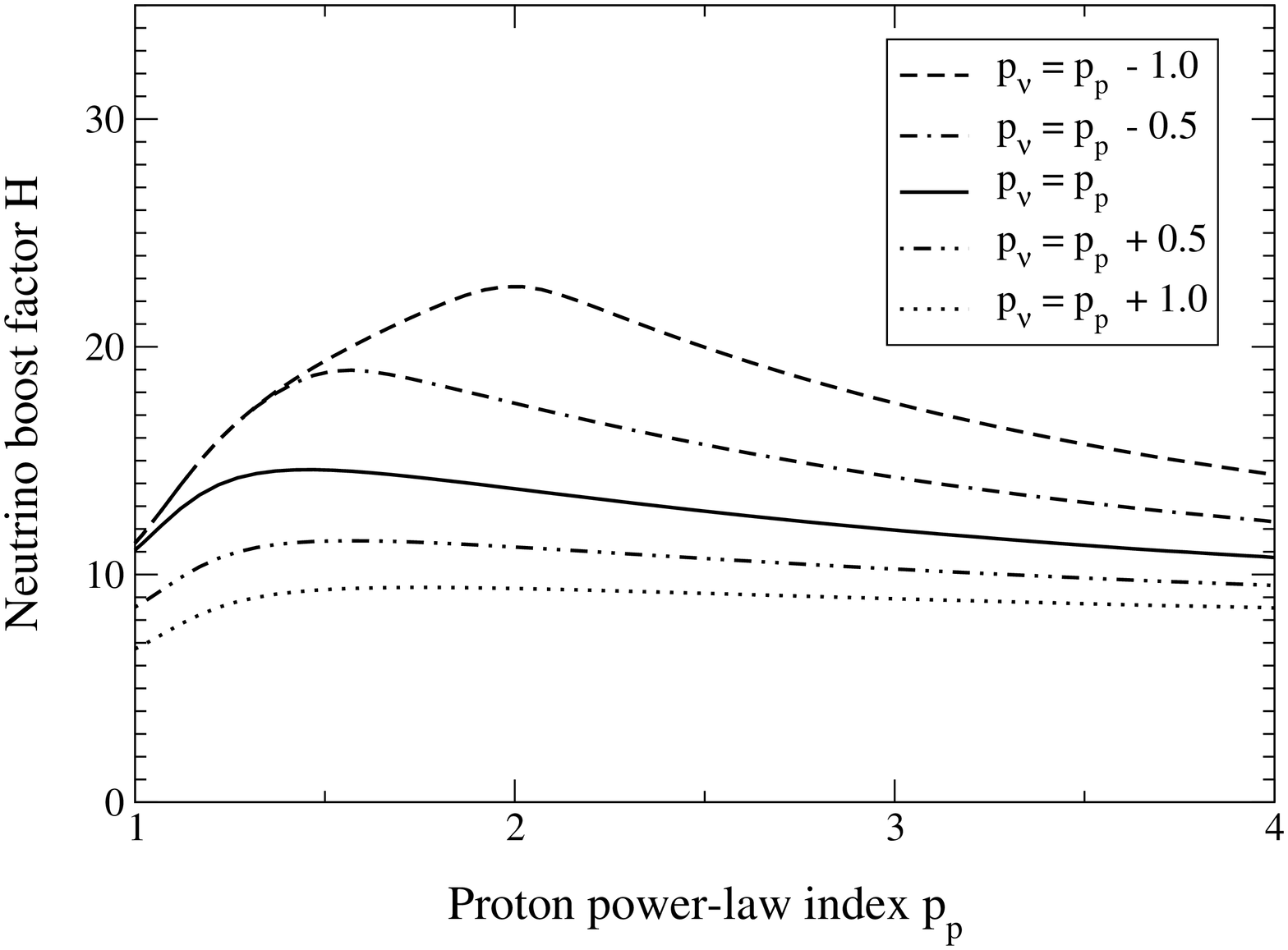}
\hspace{1cm}
\includegraphics[width=5cm, angle=270]{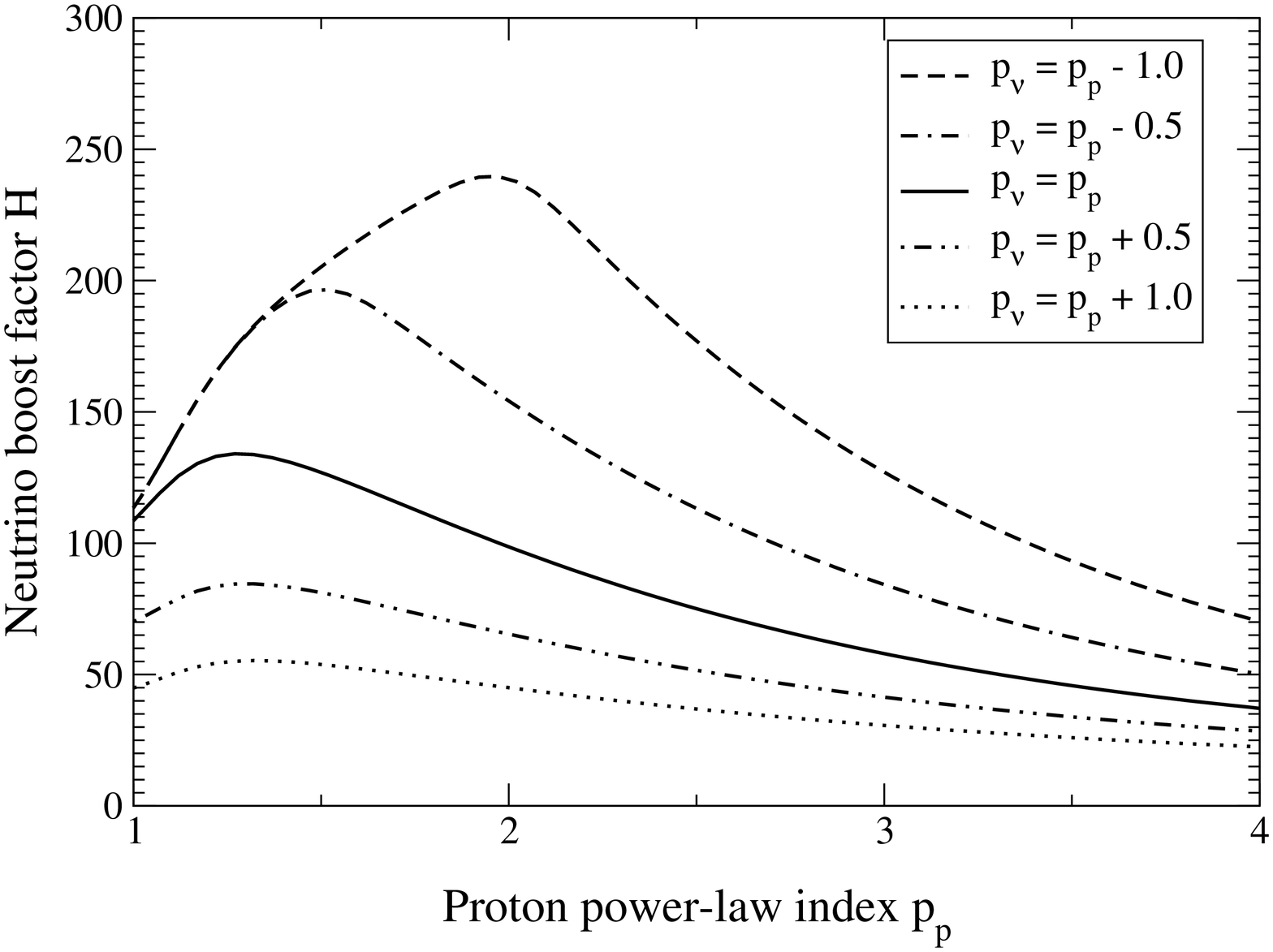} 
\caption{\label{fig:H} Neutrino boost factor $H$ as a function of
proton power-law index $p_p$ for five neutrino power-law indices
$p_\nu$ and for $E_{\rm th}$ = 57 EeV. From top to bottom the neutrino spectrum becomes progressively
softer compared to the proton spectrum:
$p_\nu = p_p -1.0$ (dashed line),
$p_\nu = p_p -0.5$ (dash-dash-dotted),
$p_\nu = p_p$ (solid),
$p_\nu = p_p + 0.5$ (dot-dot-dashed), and
$p_\nu = p_p + 1.0$ (dotted).
In the left panel we assumed no source evolution; in the right
panel we assumed that the sources follow  AGN evolution.
}
\end{figure}

In figure \ref{fig:H} we show the 
neutrino boost factor $H$ as a function of
proton power-law index $p_p$ and for  different neutrino power-law indices $p_\nu$.
In producing this figure, we 
have taken the detector threshold energy equal to $E_{\rm th}$ = 57 EeV (the energy
threshold used in the  PAO analysis \cite{Cronin:2007zz, Abraham:2007si}), and we
have assumed that the maximum proton energy is much larger than this.
As can be seen in the figure, the boost factor increases mildly as the
neutrino injection spectrum becomes harder compared to the proton injection
spectrum, which can be understood from the scaling of energy spectra with redshift.
Including source evolution is a more dramatic effect, increasing 
$H$ by an order of magnitude. This of course reflects the fact that,
with strong source evolution, the fraction of sources that can be
seen in high-energy neutrinos but not in UHECRs increases substantially.

 The systematic uncertainty of $\sim$$20\%$ in 
energy determination by  PAO \cite{Dawson:2007di}
introduces some uncertainty in
our results on the neutrino boost factor $H$.
If the actual threshold energy is lower than 57 EeV, the effect
of UHECR flux attenuation is less severe and
hence the neutrino boost factor $H$ is smaller than the
results presented in figure \ref{fig:H}. We have found that,
for the parameters used in figure \ref{fig:H}, this effect
is well approximated by the simple formula
\beqs
H(E_{\rm th}) = 10^{\frac{E_{\rm th}}{E_0}-1} H (E_0) \, ,
\eeqs
where $E_{\rm th}$ now denotes the actual threshold energy
and $E_0 \equiv 57$ EeV. As this equation shows,
the neutrino boost factor $H$ becomes
smaller (larger) by a factor $\simeq$$1.6$
in case the energy is systematically
overestimated (underestimated) by $20\%$.

\subsection{Scaling the neutrino flux from Cen A}

We now specialize the discussion to the recent  PAO data.
The number of events with energy above $E_{\rm th}$ from a point source at
declination $\delta_s$ can be expressed as
$N^{\rm pt} = \Phi^{\rm pt} (E_{\rm th}) T A(\delta_s)$, where $\Phi^{\rm pt} (E_{\rm th})$ is the
integral flux above $E_{\rm th}$, $T$ is the observation time, and $A (\delta_s)$ denotes the
experiment's effective area for a source at declination $\delta_s$. The number of events
due to the diffuse flux is then $N^{\rm diff} = \Phi^{\rm diff} \Xi $, where
$\Phi^{\rm diff}$ is the integral diffuse flux per sterad,
and $\Xi \equiv T \int A(\delta_s) d \Omega$ denotes the exposure.
With these expressions, we estimate the diffuse UHECR flux $\Phi_p^{\rm diff}$ and the UHECR
flux from Cen A $ \Phi_{p}^{\rm Cen \, A} $ 
above $E_{\rm th} = 57$ EeV to be:
\beqs
\Phi_p^{\rm diff} (E_{\rm th}) &=& \frac{N_{\rm tot} - N_{\rm Cen \, A} }{\Xi} 
= \EE{9}{-21}  {\rm cm}^{-2}\, {\rm s}^{-1} {\rm sr}^{-1} \, ; \\
\label{eq:UHECRfluxCenA}
\Phi_p^{\rm Cen \, A} (E_{\rm th}) &=& \frac{N_{\rm Cen \, A} }{\Xi } \frac{ \int A(\delta_s) d \Omega }{A(\delta_s) } 
= \EE{5}{-21}  {\rm cm}^{-2}\, {\rm s}^{-1} \, ,
\eeqs
where  $\Xi = 9000$ km$^2$ yr s is the total Auger exposure,
$\delta_s = -43 \degree$ is the declination of Cen A,
 $N_{\rm tot} = 27$ is the total number of observed UHECR events,
and $N_{\rm Cen \, A}$ is the number of events from Cen A.
Following the PAO analysis \cite{Cronin:2007zz, Abraham:2007si}
we attribute $N_{\rm Cen \, A} =2 $ events to Cen A (note however that
the actual number may be larger \cite{Moskalenko:2008iz}).
In eq. \eqref{eq:UHECRfluxCenA} we
used the relation $A(\delta_s) \propto \omega(\delta_s)$, where  $\omega$ is the
relative PAO exposure given in Ref. \cite{Sommers:2000us}, to estimate
$A(\delta_s)  / \int A(\delta_s) d \Omega = 0.15$ sr$^{-1}$.

Using the definition  \eqref{eq:def:Eta} of the neutrino boost factor $H$,
we now express the ratio of the neutrino flux from Cen A to the associated
diffuse flux as follows:
\beqs
\label{eq:fluxratio}
\frac{\phi_\nu^{\rm diff} (E)}{\phi_\nu^{\rm Cen \, A} (E)} 
= \frac{H (E_{\rm th}) \Phi_p^{\rm diff} (E_{\rm th})}{\Phi_p^{\rm Cen \, A} (E_{\rm th})} 
= \frac{H (E_{\rm th}) \lb N_{\rm tot} - N_{\rm Cen \, A} \rb A (\delta_s)}{N_{\rm Cen \, A} \int A (\delta_s) d \Omega }
= 1.9 \,  H (E_{\rm th}) \, {\rm sr}^{-1} \, ,
\eeqs
where we have used the fact that protons from Cen A reach Earth virtually without energy loss.
This equation is the main result of this paper.
Together with the results presented in figure \ref{fig:H},
it allows to estimate the diffuse neutrino flux from a model
neutrino flux for Cen A, under the assumption that the physical
environment in Cen A is representative for all UHECR and
neutrino sources. 
Within the parameter range shown in figure \ref{fig:H},
we thus find that the all-sky diffuse neutrino flux is
$\sim$$200-500$ ($800-5000$) times larger than the neutrino flux from Cen A
in the case of no (strong) source evolution.

We now compare the expected event rate in a neutrino detector for neutrinos
from Cen A to the event rate for the diffuse neutrino flux. 
In this work we focus on neutrino detection with IceCube for definiteness.
Neutrino detection with IceCube is discussed in appendix \ref{app:icecube}.
Neutrinos from Cen A are downgoing
for IceCube. Hence, for a fair comparison between Cen A and the diffuse flux,
we consider both upgoing and downgoing diffuse neutrinos
although detection of the latter is complicated by the atmospheric muon background.
From eqs. \eqref{eq:fluxratio},
\eqref{detNcenA}, and \eqref{detNdiff:dn} we find that the ratio of events 
associated with the diffuse downgoing neutrino flux to events associated with Cen A
is
\beqs
\label{eq:Nratio:dn}
\frac{N_\nu^{\rm diff, \, dn}}{N_\nu^{\rm Cen \, A}} 
\simeq 11  H (E_{\rm th})   \, ,
\eeqs
where we took the IceCube field-of-view equal to $\Omega_I = 5.7$ (up to $5\degree$
from the horizon).
In this expression we approximated the effective area for downgoing neutrinos as angle-independent.
The ratio of events 
associated with the diffuse upgoing neutrino flux to events associated with Cen A is
\beqs
\label{eq:Nratio:up}
\frac{N_\nu^{\rm diff, \, up}}{N_\nu^{\rm Cen \, A}} 
\simeq 11 \chi  H (E_{\rm th})   \, ,
\eeqs
where (cf. eqs. \eqref{detNcenA} and \eqref{detNdiff:up})
\beqs
\label{eq:def:chi}
\chi := \frac{\int d E \, \phi_\nu (E) A_{\nu, \rm eff}^{\rm up} (E) }
{\int d E \, \phi_\nu (E) A_{\nu, \rm eff}^{\rm dn} (E)}
\eeqs
is a factor of order unity that accounts for the
angular dependence of the detector.
Here  $A_{\nu, \rm eff}^{\rm dn}$ ($A_{\nu, \rm eff}^{\rm up} $) denotes
the (average) effective area for downgoing (upgoing) diffuse neutrinos.
Using estimates for the effective areas of the IceCube detector
given in eqs. \eqref{eq:Aeffnu:dn} and \eqref{eq:Aeffnu:up}, respectively,
we have verified that $1 \lesssim \chi \lesssim 2$ for neutrino test spectra $\phi_\nu \propto
E_\nu^{-p}$, where $1 \leq p \leq 3$ and \mbox{$10^3$ GeV $ < E_\nu < 10^8$ GeV}.

\section{Source model example}
\label{sec:srcex}
Following the hypothesis that 2 of the 27 UHECR
events detected by  PAO are from Cen A, different
models for ultra-high-energy (UHE) proton and neutrino
production in Cen A were proposed in Refs.
\cite{Cuoco:2007qd, Kachelriess:2008qx}.
For reasons of space we focus on the model
adopted in Ref.  \cite{Cuoco:2007qd}, hereafter referred to as the CH model,
which is based on earlier
work in Ref. \cite{Mannheim:1998wp}.
For this model, we estimate the associated diffuse neutrino flux and compare
its detection prospects to those of the neutrinos from Cen A using
the general results obtained in the previous section.
In the last part of this section we comment on the detection prospects
of the models that were recently proposed in Ref.
\cite{Kachelriess:2008qx}.

\subsection{The model}
An attractive feature of the model adopted in Ref. \cite{Cuoco:2007qd}
is that the neutrino flux
at high energies is harder than the flux of UHECR protons.
To achieve this, the model requires that a population of high-energy seed protons
(accelerated through e.g. shock acceleration)
is confined to a region close to the source.
These protons create neutrons and neutrinos in photopion interactions
with the ambient photon field. The neutrons  escape from the
source, decay, and give rise to UHECR protons. In this process
the neutrons lose energy in interactions with the photon field before decay,
thus softening the spectrum of UHE protons.
Neutrinos, on the other hand, trace the neutron energy spectrum upon production
(i.e. without energy loss).

The model predicts two spectral breaks in the cosmic-ray injection spectrum
at the energies where the optical depths for proton and neutron 
photopion production become unity
(see Ref. \cite{Mannheim:1998wp} for a thorough discussion). Since these
energies are generally similar, one may assume a single break energy $E_{\rm br}$.
Below this break energy both the cosmic-ray injection spectrum and the neutrino spectrum
are harder than the seed proton spectrum by one power of the energy (assuming
photopion production on a $n_\gamma (\ep_\gamma) \propto \ep_\gamma^{-2}$ photon field).
Above the break energy the  cosmic-ray injection spectrum is softer than
the seed proton spectrum by one power of the energy (assuming the same photon field),
while the neutrino
spectrum follows the initial seed proton spectrum.
Identifying the high-energy part of the injected cosmic-ray spectrum with
the observed UHECR flux $\phi_p$, we can express the all-flavor neutrino flux $\phi_{\nu_x}$ as:
\beqs
\label{eq:nuflux}
\phi_{\nu_x} (E) = \frac{\xi_\nu}{\xi_n \eta_{\nu n}^2}
{\rm min} \lb  \frac{E}{\eta_{\nu n} E_{\rm br}} ,  \frac{E^2}{\eta^2_{\nu n} E^2_{\rm br}}\rb
\phi_p \lb  \frac{E}{\eta_{\nu n}} \rb  \, ,
\eeqs
where  $\xi_i$ is the energy fraction of the
proton that is transferred to species $i$ (neutron or neutrino) in photopion interactions,
and $\eta_{\nu n}$ is the ratio of
the average neutrino energy to the average neutron energy. The values of these
quantities depend on the spectral distribution of the photon field and can be estimated numerically.
The authors of Ref.  \cite{Cuoco:2007qd} take $\xi_\nu / \xi_n = 0.2$ and
$\eta_{\nu n} = 0.04$ for interactions in the nucleus of Cen A.
The break energy, being due to a change in photopion production efficiency,
is determined by the ambient photon distribution. Although its value cannot be directly
inferred from observations, it may be estimated from the observed gamma-ray spectrum because
interactions  between gamma rays and the ambient photons also give rise
to a spectral break  in the gamma-ray flux. In this way the break energy in the
UHECR spectrum can be
estimated as
$E_{\rm br} \simeq \EE{3}{8} \, E_{\gamma, \rm br}$,
where  $E_{\gamma, \rm br}$ denotes the gamma-ray break energy
\cite{Mannheim:1998wp, Cuoco:2007qd}.
The authors of Ref. \cite{Cuoco:2007qd}
conservatively take 
$E_{\gamma, \rm br} \simeq 200 $ MeV for Cen A, so that 
$E_{\rm br} \simeq 10^8$ GeV.

 In this work we only consider detection of muon neutrinos for definiteness,
and hence we should take account of  the neutrino flavor ratios.
In the CH model, the source is optically thin and the effect of
meson synchrotron energy loss is neglected. Under these conditions,
the neutrino flux from proton-photon interactions 
is dominated by neutrinos from pion decay at all energies \cite{Kachelriess:2006fi}. 
This implies that the neutrino flavor ratio (electron : muon : tau) at the source is
approximately $1:2:0$, as is also indicated in fig. 1 of Ref. \cite{Cuoco:2007qd}.
In this case neutrino oscillations  lead to a flavor ratio close to $1:1:1$
at Earth \cite{Learned:1994wg}.
Hence the expected muon-neutrino flux from Cen A, in the CH model, is:
\beqs
\label{eq:nuflux:CenA}
\nn \phi_{\nu}^{\rm Cen \, A} (E)
&=& \frac{\Phi_p^{\rm Cen \, A} (E_{\rm th})}{3} \frac{\xi_\nu \eta_{\nu n}^{p-2}}{\xi_n}\frac{(p-1) }{ E_{\rm th}} \lb \frac{E}{E_{\rm th}} \rb^{-p} 
\lb \frac{E}{E_{\nu, \, \rm br}} \rb  {\rm min} \lb 1,  \frac{E}{ E_{\nu, \, \rm br}} \rb \\
& =  &  \EE{3}{-11} \lb \frac{E}{1 \, {\rm GeV}} \rb^{-1.7}
{\rm min} \lb 1,  \frac{E}{ E_{\nu, \, \rm br}} \rb
 \, {\rm GeV}^{-1}\,{\rm cm}^{-2}\, {\rm s}^{-1}
\, ,
\eeqs
where $E_{\nu, \, \rm br} \equiv \eta_{\nu n} E_{\rm br} = \EE{4}{6}$ GeV.

\begin{figure}
\includegraphics[width=5cm, angle=270]{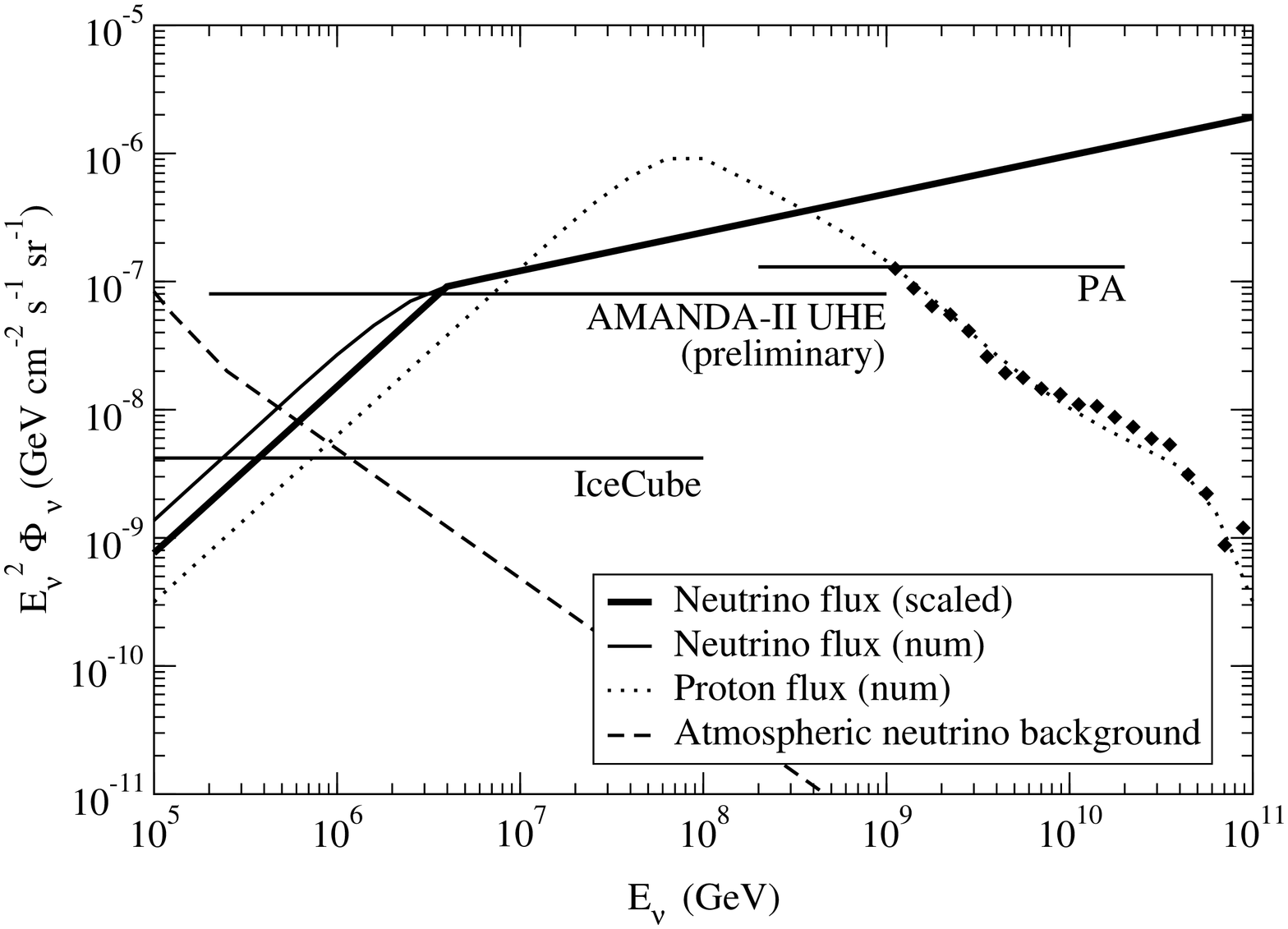}
\hspace{1cm}
\includegraphics[width=5cm, angle=270]{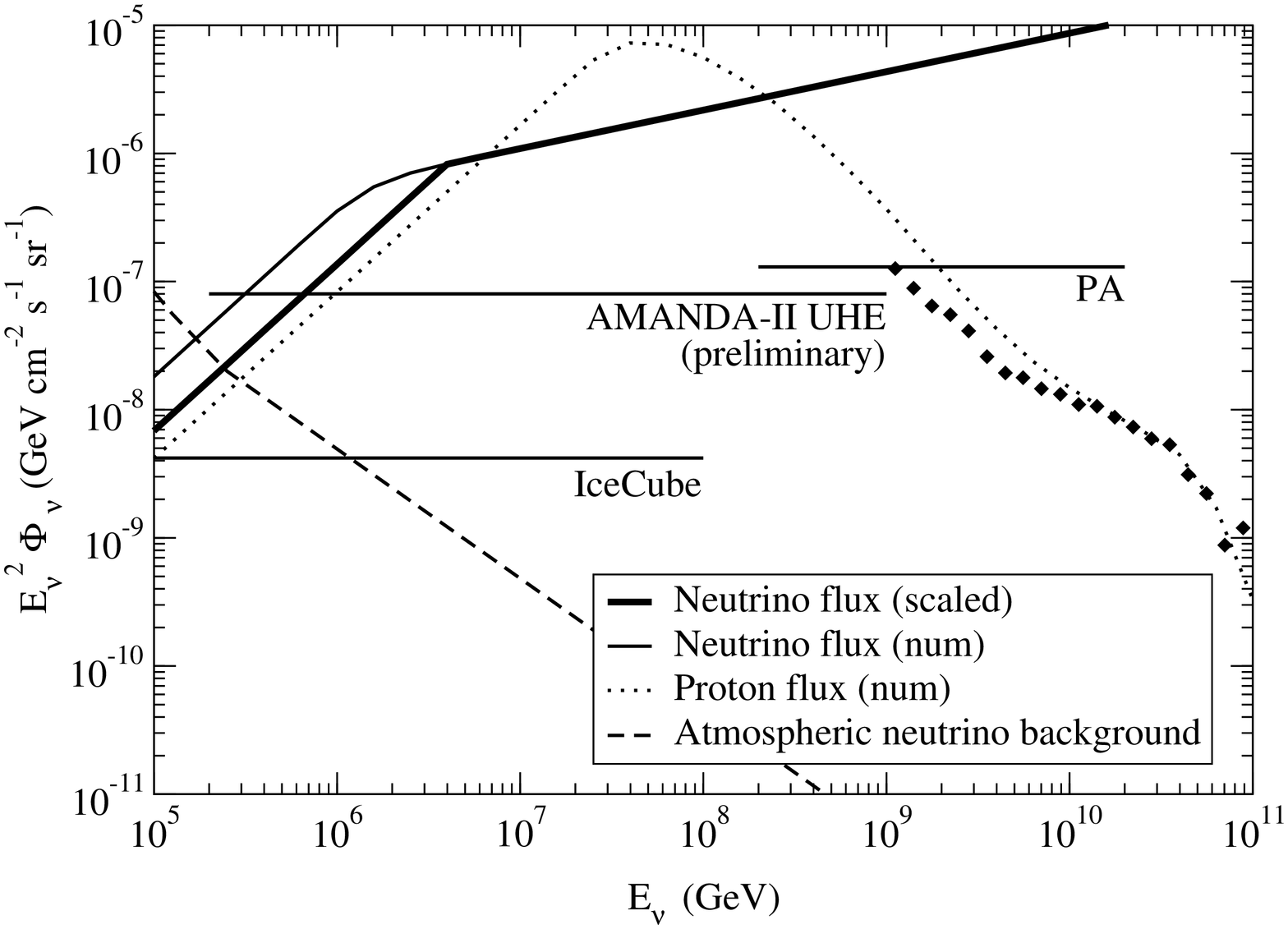} 
\caption{\label{fig:fluxes} 
Diffuse muon-neutrino flux obtained by scaling (thick solid line)
and by numerical computation (solid), together with the diffuse
UHECR flux obtained numerically (dotted) and the
atmospheric muon-neutrino background  (dashed) as a function of 
neutrino energy. Also shown are existing upper limits from AMANDA-II
and  PAO, the projected upper limit for IceCube (3 years), and
UHECR data from  PAO (diamonds). Left panel: no source evolution;
right panel: strong source evolution.}
\end{figure}

\subsection{Diffuse neutrino flux}
The neutrino boost factor for a proton power-law spectrum with index $p=2.7$ and
a neutrino power-law spectrum with index $p=1.7$,
with and without source evolution, is $H  = 19 $ and $H = 159$,
respectively (see fig. \ref{fig:H}). Scaling
the model neutrino flux from Cen A from eq. \eqref{eq:nuflux:CenA}, we
find the following estimates for the diffuse muon-neutrino flux within the CH model:
\beqs
\label{eq:nuflux:diff}
\phi_{\nu}^{\rm diff} (E) \simeq
10^{-9} \lb \frac{E}{1 \, {\rm GeV}} \rb^{-1.7}\, {\rm min} \lb 1, \frac{E}{ E_{\nu, \, \rm br}} \rb \, {\rm GeV}^{-1}\, {\rm cm}^{-2}\, {\rm s}^{-1} {\rm sr}^{-1} \, 
\eeqs
in the case of no source evolution, and a factor of 10 larger in the case
of strong evolution.
These fluxes are shown in figure \ref{fig:fluxes}. Also shown are
numerical results (obtained by the method described in section \ref{sec:att}) for the diffuse UHECR 
and neutrino fluxes after propagation,  the preliminary 
AMANDA-II UHE limit \cite{Gerhardt:2007zz}, the  PAO limit \cite{Abraham:2007rj},
the projected limit for IceCube \cite{Ahrens:2003ix} after three years of data-taking,  PAO data on the UHECR proton flux 
\cite{Yamamoto:2007xj}, and the atmospheric neutrino background.
The AMANDA-II,  PAO, and IceCube detection limits are the  90\% confidence level
upper limits for a $\phi_\nu \propto E_\nu^{-2}$ diffuse muon-neutrino flux,
where we have assumed a $1:1:1$ flavor ratio at Earth.
The atmospheric muon-neutrino background flux is parameterized as follows:
\beqs
\label{eq:atmbg}
\phi_\nu^{\rm bg}(E_\nu)  
 = \left\{ 
\begin{array}{l r}
\frac{\EE{8.4}{-2} \lb E_\nu/1 \, {\rm GeV} \rb^{-2.74} }
{1 + 0.002 \lb E_\nu / 1\,  {\rm GeV} \rb } \, {\rm GeV}^{-1} \,  {\rm cm}^{-2} \, {\rm s}^{-1} \, {\rm sr}^{-1}  &
(E_\nu < 10^{5.3} \, {\rm GeV}) \\
\EE{5.7}{-3} \lb E_\nu / 1 \,  {\rm GeV} \rb^{-3.01} 
\, \, {\rm GeV}^{-1} \,  {\rm cm}^{-2} \, {\rm s}^{-1} \, {\rm sr}^{-1} & (E_\nu > 10^{5.3} \, {\rm GeV})
\end{array}
\right.
\, ,
\eeqs
where the high-energy contribution 
is due to prompt charm decay. This parameterization is, by construction,
close to the maximum background indicated in figure 11 of 
Ref. \cite{Ahrens:2003ix}.

We observe from figure \ref{fig:fluxes} that the scaled neutrino flux is an excellent approximation to
the numerical results in the high-energy regime, where the energy spectrum follows
a single power-law. The spectral break in the scaled neutrino flux is however much sharper
than in the numerical results,
where it is smoothened due to an averaging over redshift. As a result,
the approximation obtained by scaling underestimates
the resulting diffuse neutrino flux at lower energies. 
We have verified that this does not strongly
affect the expected number of neutrino events in IceCube.
(In principle, the diffuse flux at energies below and above
the break can be estimated independently from the results obtained
is section \ref{sec:att}, which would give
a better estimate. Given the other uncertainties we are faced
with, we will not pursue this though.)
In producing this figure, we have assumed that the maximum neutrino
energy is larger than $10^{11}$ GeV. The exact value is not very important
for our estimates, as the interaction rate in IceCube is dominated
by neutrinos of energy below $10^8$ GeV.

We note from the figure that the UHECR flux follows the  PAO data
above 57 EeV, which is a consistency check of our numerical method.
The model flux for strong evolution is marginally consistent
with the  PAO data at lower energies, given the large uncertainties in
energy calibration.

The estimated diffuse neutrino flux for the CH model is well
above AMANDA-II and  PAO limits in the case of strong evolution.
This implies that either the CH model (when applied to strongly evolving
sources) is too optimistic,
Cen A is intrinsically exceptional, or that  significantly more than two of the
UHECR events observed by  PAO are from Cen A.
In the case of no evolution, the diffuse flux is only
marginally above these limits. (Note that the limits are based on
a  neutrino flux $\phi_\nu \propto E^{-2}$ , so that a more accurate comparison requires further
analysis.)
The flux is however well
above the projected IceCube upper limit. At energies larger than $\sim$$10^6$ GeV,
the atmospheric neutrino background is strongly suppressed 
and hence detection should pose no difficulties.

\subsection{Event rates}
Using estimates for the neutrino effective area presented
in appendix  \ref{app:icecube} (see fig. \ref{fig:Aeff}), we
can now estimate the neutrino event rate of the diffuse neutrino flux and that
from Cen A in the IceCube neutrino detector.
Since Cen A is in the southern hemisphere,
neutrinos from this galaxy are downgoing for IceCube. This makes detection
challenging, and only possible at very high energies.
To eliminate the background we consider the flux of
neutrinos with energy
\mbox{$10^6$ GeV $ < E_\nu < 10^8$ GeV}
(cf. fig. \ref{fig:fluxes}).
For a fair comparison between the neutrino interaction rates for Cen A and
the diffuse flux, we consider both up- and downgoing diffuse neutrinos.
We find that  $\chi = 1.4$ for the CH model, where  $\chi$ is the factor 
that enters the scaling relation
eq. \eqref{eq:Nratio:up}.
We adopt a field of view $\Omega_I = 5.7$,
corresponding to cutting at $5\degree$ below or above the horizon.

 From eq. \eqref{detNcenA},
we find $0.08$ events per year for Cen A with an expected background
of $0.004 \, (\theta/5\degree)^2$, where $\theta$ is the angular resolution.
This is in reasonable agreement with results obtained
in Ref. \cite{Cuoco:2007qd}, where an event rate of $0.35$ per year is found for
neutrinos of all flavors (assuming equal detection probabilities).
Using the scaling relations eqs. \eqref{eq:Nratio:dn} and \eqref{eq:Nratio:up},
we directly obtain our estimate of 17 (24) neutrino events per year due to
downgoing (upgoing) diffuse neutrinos per year in the case if no source evolution.
Including source evolution, we find 145 (203)  downgoing (upgoing) events per year.
In this energy range the number of background events for the
downgoing (upgoing) diffuse neutrino flux is roughly 0.4 (1) per year,
and hence detection is virtually background-free.

Our results for the event rate due to
the diffuse flux of upgoing neutrinos  in the CH model  are larger than
recent estimates in Ref. \cite{Halzen:2008vz}, who estimate $\sim$$5$ events per year
for this model. 
This result is to be compared with our no-evolution estimate of 24 events per year.
The authors of Ref. \cite{Halzen:2008vz} attribute the diffuse neutrino and UHECR flux
to Fanaroff-Riley I (FRI) radio galaxies, of which Cen A is an example.
They estimate the diffuse neutrino flux by adding the contribution of sources up to
$z=0.5$ using the inferred FRI source density.
In contrast, our estimates are normalized to the UHECR flux and we
consider sources up to $z=5$.  It is reassuring that these results, 
which are obtained in a different manner,  are within a factor
few.
The fact that our estimates are 
somewhat larger
may be attributed to the larger maximum redshift that we have adopted
(see also the comment at the end of section III of Ref. \cite{Halzen:2008vz}).

\subsection{Comparison with other models}
In the above we have focused on a model for Cen A that was 
put forward in Ref. \cite{Cuoco:2007qd}. More recently,
the authors of Ref. \cite{Kachelriess:2008qx} have also discussed 
UHE proton and neutrino production in Cen A.
The authors consider three different spectra for the accelerated protons:
(i) a straight power-law spectrum with index $p=2.0$,
(ii) a broken power-law spectrum with $p=2.0$ before and $p=2.7$ after the break energy, 
and (iii) a straight power-law spectrum with index $p=1.2$. The first
two spectra may be the result of stochastic shock acceleration, the last spectrum
of linear acceleration in a regular electric field.
In case of acceleration near the core neutrinos are produced 
predominantly in the interaction of the accelerated protons with UV photons
while low-energy protons provide the dominant
target for neutrino production if the protons are accelerated in the jet \cite{Kachelriess:2008qx}. 
The authors obtain neutrino spectra for the three acceleration mechanisms,
applied to both the core and the jets, numerically. 
We do not attempt to reproduce these here, but rather estimate the associated
diffuse neutrino flux and the event rates from their results.

For the three acceleration models (i), (ii)
and (iii), we find that $H = 14$, $13$, and $14$, respectively, in the
case of no evolution (see also fig. \ref{fig:H}). 
Including source evolution, we find $H=101$,
$69$, and $136$, respectively.
In deriving these estimates we have approximated the
neutrino spectra as tracing the proton spectra; for
softer spectra the results are somewhat lower.
Scaling the neutrino fluxes presented in fig. 1 of Ref. \cite{Kachelriess:2008qx}
with these values,
we find that the diffuse neutrino fluxes corresponding to model (ii)
are well above the existing AMANDA-II limits \cite{Achterberg:2007qp, Gerhardt:2007zz}
in the case of strong source evolution.
For all models except the linear accelerator (iii) the expected event
rates in IceCube (obtained by scaling the results presented in table
1 of Ref. \cite{Kachelriess:2008qx}) are $\gg 1$ per year,
even in the case of no source evolution.
Hence IceCube should be able to put strong constraints on these
models.

\section{Summary and discussion}
\label{sec:conclusion}
In this work we have considered the relation between the diffuse neutrino
flux and the neutrino flux from Cen A under the hypothesis that Cen A is
a characteristic source of UHECR protons and UHECR-related neutrinos. This is motivated by 
recent results from  PAO
\cite{Cronin:2007zz, Abraham:2007si} which suggest
that Cen A may be a source of UHECRs. If it is also a source of
neutrinos, as proposed in Refs. \cite{Cuoco:2007qd, Kachelriess:2008qx},
and if the environment in Cen A is representative for other sources,
we argue that the diffuse neutrino flux may also constrain the source
model.  The diffuse neutrino
flux can easily be estimated by scaling a model Cen-A neutrino flux
(see eq. \eqref{eq:fluxratio}) 
if one assumes that Cen A is a `typical' source, i.e. under the assumption
of universal UHECR and neutrino injection spectra with a fixed relative strength.
We stress that we make no assumptions on the intrinsic
luminosity or distance of the sources.
We have derived estimates for
the corresponding scaling factor in section \ref{sec:att}
without relying on a particular source model (see fig.  \ref{fig:H}). The scaling factor depends
mildly on both the assumed proton and neutrino energy spectra, but very strongly
on the assumed model of source evolution.
This suggests that constraints related to the diffuse neutrino flux
may be especially useful in constraining the evolution of UHECR sources.

Regardless of the source model, we find that the estimated neutrino event rate in IceCube due to the
diffuse neutrino flux is expected to be at least two orders of magnitude 
larger than the event rate of neutrinos from Cen A, as may be seen
from eqs. \eqref{eq:Nratio:dn} and \eqref{eq:Nratio:up}. When sources follow strong AGN evolution as parameterized
in Ref. \cite{1998MNRAS.293L..49B}, the rate is even three orders of
magnitude higher. Therefore we conclude that the detection of neutrinos
with IceCube, without the detection of the associated diffuse neutrino
flux, would imply that Cen A is an exceptionally efficient neutrino source.
Here we assume that the neutrino flux extends to energies above $10^6$ GeV,
so that neutrino detection is not limited by background rejection.

We have applied our results to models recently proposed in the literature
in section  \ref{sec:srcex}.
We find that the diffuse neutrino flux associated with the model
adopted in Ref. \cite{Cuoco:2007qd} is well above the (preliminary)
AMANDA-II UHE limit \cite{Gerhardt:2007zz} and the  PAO limit \cite{Abraham:2007rj}
when the sources follow strong evolution.  This
implies that either the model (when applied to strongly evolving sources)
overpredicts the neutrino flux,
Cen A is intrinsically
exceptional, or that considerably more than two of the UHECR events observed by
PAO are produced by Cen A.
Similarly, we find that the diffuse flux associated with the most optimistic models
considered in Ref. \cite{Kachelriess:2008qx} 
is also above AMANDA-II limits   \cite{Achterberg:2007qp, Gerhardt:2007zz} for strong source
evolution. The
expected event rate in IceCube is much larger than 1 yr$^{-1}$ for all models considered in 
Ref. \cite{Kachelriess:2008qx} except for the linear accelerator (both in the case
of no source evolution and of strong source evolution). IceCube should thus be able to put strong constraints on these
models.

 Several comments are in order.
First of all, it is presently unknown to which extent UHECR source are similar in nature,
and whether or not Cen A is a typical cosmic-ray source
(recall that we use `typical' only as
a statement on the ratio of outputted neutrino flux to cosmic-ray flux;
it bears no meaning to the intrinsic luminosity
or the distance).
We have found no a priori reason
to suppose that Cen A is an atypical cosmic-ray source.
Cen A is quite representative of the local
population of radio-loud AGNs \cite{Moskalenko:2008iz},
which, although constituting a
subdominant fraction of $15-20\%$ of all AGNs \cite{Urry:1995mg},
are commonly (though not uniquely) considered as the dominant
cosmic-ray sources because of their powerful jets.
Notwithstanding this plausibility argument, the question whether or
not Cen A is indeed 
a typical UHECR source can 
only be resolved by further observations.
Our results address this question by placing constraints on the
simplest scenario, namely that Cen A is a typical source, representative of a universal
class of cosmic-ray sources. In this context we reiterate our conclusion
that the detection of neutrinos from Cen A, without the detection of the diffuse
neutrino flux, would rule out Cen A as a typical cosmic-ray source
on the ground of its exceptionally high neutrino production
efficiency.

A second comment regards the number of UHECR events attributed to Cen A.
In our numerical estimates we have
followed the PAO analysis, which associates two events with Cen A as
the angles between their
arrival directions and the nucleus of Cen A are smaller than $3.1\degree$. 
It has been pointed out in Ref. \cite{Moskalenko:2008iz}
that at least four events can be associated with Cen A if one takes account
of its extended structure (the radio lobes subtend $\sim$$9\degree$
on the sky). On the other hand, it is also possible that the observed UHECR events
are not from Cen A but rather from sources in the more distant Centaurus 
supercluster, as suggested in Ref. \cite{Ghisellini:2008gb}. 
In both cases our estimates on the neutrino boost factor
would be reduced by a factor $\sim$$2$. This would not 
qualitatively affect the conclusions for the models studied in this work.
Future data from PAO will
increase the rather limited statistics and  will very likely shed
more light on the UHECR production rate of Cen A.

Thirdly, the relation between the produced neutrino and UHECR fluxes
may be complicated when neutrinos and protons are emitted within
cones of different opening angle
(as in the model discussed in Ref. \cite{Becker:2008nf}),
or when the source luminosity varies in time.
These effects will average out for diffuse fluxes but may affect the emission
from Cen A. Any collimation in the direction
of the jet will reduce the visibility on Earth as Cen A is viewed off-axis.
Because the opening angle of the neutrino emission cone is expected to be smaller than that
of the proton emission cone, this further decreases the expected neutrino
flux from Cen A compared to 
the diffuse neutrino flux. Radio and X-ray measurements of Cen A
show variability on a time scale shorter than a year \cite{1998A&ARv...8..237I}. The arrival times
of protons and neutrinos produced during a flaring phase will however
not be correlated as the proton path length is increased by its motion
in the galactic magnetic field. Hence a strong flare may in principle lead to
an increased neutrino flux without an increase in the  observed
UHE proton flux from 
Cen A.

Finally, we note that the diffuse gamma-ray flux produced by sources similar
to Cen A  
may also constrain the source model, in a fashion very similar to the 
diffuse neutrino flux that was considered here. 
Gamma-ray emission by Cen A was considered in Refs.
\cite{Kachelriess:2008qx, Gupta:2008tm} but these authors have not considered
the associated diffuse gamma-ray flux.

\begin{acknowledgments}
We thank Alessandro Cuoco, Francis Halzen, Michael Kachelrie\ss, and Jelena Petrovic
for comments and discussions,  and the referee for useful and contructive remarks.
H.K. and P.T. are supported by Belgian Science Policy under IUAP VI/11
and by IISN. The work of P.T. is supported in part by the
FNRS, contract 1.5.335.08.
\end{acknowledgments}

\appendix
\section{Proton energy loss approximation}
\label{app:beta0}
Proton energy loss during propagation from a source at redshift $z$
can be described by the differential equation
(see, e.g., Ref. \cite{Berezinsky:2002nc})
\beqs
\label{eq:Eg:diffeq}
\frac{1}{E} \frac{dE}{dz} = \frac{1}{1+z} + \frac{(1+z) \beta_0 ((1+z)E)}{H (z)}  \, ,
\eeqs
where the first term accounts for redshift energy loss and the second for
particle interactions.
The function $\beta_0$ gives the inverse energy-loss time (at present epoch)
for interactions between protons and CMB photons.

In this work we determine the energy at the source $E_0$ as a function
of observed energy $E$ and redshift $z$ by solving 
eq. \eqref{eq:Eg:diffeq} numerically.
To do this, we split $\beta_0 = \beta_0^{\pi} + \beta_0^{ee}$ into
two parts corresponding to energy loss due to
pion photoproduction and electron-positron
pair production, which we approximate by:
\beqs
\log \beta_0^{\pi} &=& \sum_{n=1}^5 a_n X^{n-1} \, ;\\
\log \beta_0^{ee} &=& \sum_{n=1}^5 b_n X^{n-1} \, ,
\eeqs
where $X = \log (E / 1 \,  {\rm eV})$, and $\beta_0$ is expressed in units of yr$^{-1}$.
For $10^{10.5}$ GeV $  < E < 10^{12}$ GeV,
$\vec{a} =  \lb \EE{-1.2}{5}, \, \EE{2.3}{4},\,  \EE{-1.7}{3}, \, 52, \, -0.62\rb$;
for higher energies $a_1 = -7.6$ is the only non-zero coefficient; for
lower energies $\beta_0^{\pi}=0$.
For $10^{8.5}$ GeV $  < E < 10^{12}$ GeV,
$\vec{b} =  \lb \EE{-1.3}{4}, \, \EE{2.6}{3},\,  \EE{-2.0}{2},\,  6.6,\,  -0.082\rb$;
at lower or higher energies $\beta_0^{ee}=0$.

\section{Neutrino detection with IceCube}
\label{app:icecube}
In this section we estimate the effective area for the detection of muon-neutrinos with IceCube.
The word `neutrino' refers to muon-neutrino throughout this section.

\subsection{Neutrino interaction and muon track length}
The expected number of neutrino events in IceCube for a neutrino source with differential flux $\phi_\nu^{\rm pt}(E_\nu)$ at an angle $\theta$
with respect to the nadir (i.e., $\theta=0$ points towards the North Pole) can be written as follows:
\beqs
\label{eq:Nnupt}
N = T \int d E_\nu \,  \phi_\nu^{\rm pt} (E_\nu) \mathcal{P}_\nu (E_\nu,  \theta)
\int d E_\mu \, n(E_\nu, E_\mu) 
\mathcal{P}_\mu (E_\mu) A_{\mu, \rm eff} (E_\mu) \,  ,
\eeqs
where $T$ is the observation time; $E_\nu $ the neutrino energy;
$E_\mu $ the muon energy; $\mathcal{P}_\nu$ the probability that a neutrino
reaches the vicinity of the detector; $\mathcal{P}_\mu$ the probability
that a muon is created that reaches the detector with sufficient energy
for detection; $A_{\mu, \rm eff}$ the detector's muon effective area
(which is close to the geometrical surface for high-energy muons); and
$n(E_\nu, E_\mu) $ is the muon energy distribution resulting from the
interaction of a neutrino with energy $E_\nu$. In the following we 
evaluate these quantities and make a number of simplifying assumptions.

Neutrinos interact with Earth nuclei through both
charged- and neutral-current interactions. Interactions of the former
type lead to electron, muon, and tau production while interactions
of the latter type degrade the neutrino energy. Here we assume that
all interactions transform the neutrino and thus we neglect the `regeneration' of lower-energy neutrinos by neutral-current interactions.
The neutrino survival probability $\mathcal{P}_\nu$ can then
be expressed as
\beqs
\mathcal{P}_\nu (E_\nu, \theta) = \exp \lb - N_A \sigma_{\nu N} (E_\nu)
\int_0^L \rho(r) dl  \, \rb \, ,
\eeqs
where $N_A = \EE{6.2}{-23} \, {\rm cm}^{-3}$ water equivalent (w.e.) 
is Avogadro's constant; $\sigma_{\nu N}$ is the total neutrino-nucleus 
cross section (including charged-current and neutral-current interactions);
$\rho (r) $ is the density of the Earth
as a function of the radial coordinate $r = \sqrt{l^2 + r_E^2 - 2 l r_E \cos \theta}$; and $L = 2 r_E \cos \theta$ is 
the propagation distance for a neutrino through the Earth with nadir
angle $\theta$ ($r_E =\EE{6.4}{8} $ cm is the radius of the Earth).
Since there is no experimental data on neutrino-nucleus
interactions at the UHECR energy scale one has to rely on models
to extrapolate the data from lower energies.
In this work we use the neutrino-nucleus cross sections 
tabulated in Ref. \cite{Gandhi:1995tf}. 
We adopt the Preliminary Reference Earth Model 
(see Ref. \cite{Gandhi:1995tf}) for the density 
profile of the Earth $\rho(r)$. The muon detection probability $ \mathcal{P}_\mu$
is:
\beqs
\mathcal{P}_\mu (E_\mu) = 1 -  \exp \lb - N_A \sigma^{\rm CC}_{\nu N}
(E_\nu) R_\mu (E_\mu) \rb \, ,
\eeqs
where $ \sigma^{\rm CC}_{\nu N}$ is the charged-current cross section,
and $R_\mu$ is the muon range within which the muon energy degrades to a minimum energy $E_\mu^{\rm min}$. This range can be approximated with
\beqs
R_\mu (E_\mu) = \frac{1}{b} \ln \lb \frac{a + b \, E_\mu}{a + b \,
E_\mu^{\rm min}} \rb \, ,
\eeqs
where $a = \EE{2.0}{-3} \, {\rm GeV \, cm}^{-1}$ (w.e.) accounts for
ionization and $b = \EE{3.9}{-6} \, {\rm cm}^{-1}$ (w.e.) for
radiation losses \cite{Gandhi:1995tf}. 
In this work we adopt $E_\mu^{\rm min} = 10^2$ GeV.
For simplicity we assume that 
a neutrino interaction leads to a single muon with energy
$E_\mu = y_{\rm CC}  (E_\nu) \, E_\nu$, i.e.
\beqs
n (E_\nu, E_\mu) = \delta ( E_\mu -  y_{\rm CC}  (E_\nu) \, E_\nu) \, ,
\eeqs
where the charged-current inelasticity $y_{\rm CC}  $ is tabulated in
Ref. \cite{Gandhi:1995tf}. Lastly, we use the IceCube muon effective area given
in Ref. \cite{Ahrens:2003ix}. This is the only quantity in our estimates that 
accounts for detector efficiency.

\subsection{Effective area for downgoing neutrinos}
Since downgoing neutrinos reach the detector virtually unhindered, we may approximate
$\mathcal{P}_\nu \simeq 1$ and $\mathcal{P}_\mu  \simeq N_A \sigma^{\rm CC}_{\nu N}
(E_\nu) \, {\rm min} \lb R_\mu, R_d \rb $, where $R_d \simeq \EE{2}{5}$ cm denotes the
detector depth.  At the energies where downgoing muons are detectable the interaction length
is determined by the detector depth (i.e., $R_d < R_\mu$). With these simplifications,
we express the expected number of neutrino interactions for a single point source above IceCube as
\beqs
\label{detNcenA}
N^{\rm pt, \, dn} = T \int d E_\nu \, \phi^{\rm pt}_\nu (E_\nu)   A_{\nu, \rm eff}^{\rm dn}  (E_\nu) \, ,
\eeqs
where $T$ is the observation time, $ \phi^{\rm pt}_\nu$ is the neutrino flux from the source
and  $ A_{\nu, \rm eff}^{\rm dn}$  denotes the neutrino effective area for downgoing neutrinos:
\beqs
\label{eq:Aeffnu:dn}
A_{\nu, \rm eff}^{\rm dn} (E_\nu)  = T R_d N_A  \sigma^{\rm CC}_{\nu N} (E_\nu)  A_{\mu, \rm eff}(E_\mu) \, .
\eeqs
For simplicity we take the muon effective area for downgoing neutrinos equal to the
geometrical area, $A_{\mu, \rm eff} =10^{10}$ 
cm$^2$, so that the corresponding neutrino effective area is angle-independent.
In this case we may express the number of events due to the diffuse flux of downgoing neutrinos
as follows:
\beqs
\label{detNdiff:dn}
N^{\rm diff, \, dn} = T \Omega_{\rm I} \int d E_\nu  \, \phi_\nu^{\rm diff} (E_\nu) A_{\nu, \rm eff}^{\rm dn} (E_\nu)\, ,
\eeqs
where $\Omega_{\rm I}$ is the detector's opening angle
and $\phi_\nu^{\rm diff}$ is the diffuse neutrino flux.
We take the maximum viewing angle for upgoing (downgoing) events equal to
$5\degree$ below (above) the horizon, so that $\Omega_{\rm I} = 5.7$.

We note that the detection of downgoing neutrinos
is challenging and requires special analysis techniques.
Consequently our estimates for the detection
rate of downgoing neutrinos may be too optimistic.

\subsection{Effective area for diffuse upgoing neutrinos}
For upgoing neutrinos the event rate in angle-dependent because the
neutrino survival probability depends on the incident angle.
We compute the
number of neutrino events for the diffuse flux of upgoing neutrinos 
by integrating eq. \eqref{eq:Nnupt} over the angle $\theta$. We express the
result as follows:
\beqs
\label{detNdiff:up}
N^{\rm diff, \, up} = T \Omega_{\rm I} \int d E_\nu  \, \phi_\nu^{\rm diff} (E_\nu) A_{\nu, \rm eff}^{\rm up} (E_\nu)\, .
\eeqs
Here the average effective area for diffuse up-going neutrinos,
using the simplifying assumptions described above, is given by:
\beqs
\label{eq:Aeffnu:up}
A_{\nu, \rm eff}^{\rm up} (E_\nu)  = \mathcal{S} (E_\nu)
\mathcal{P}_\mu (E_\mu) A_{\mu, \rm eff} (E_\mu) \, ,
\eeqs
where we adopt values for $A_{\mu, \rm eff} (E_\mu)$
from Ref. \cite{Ahrens:2003ix}, and 
the shadowing factor $\mathcal{S}$ is:
\beqs
\mathcal{S} (E_\nu) \equiv \frac{1}{ 1- \cos \theta_{\rm max}} \int_0^{\theta_{\rm max}} d \theta \sin \theta
\, \mathcal{P}_\nu (E_\nu, \theta) \, .
\eeqs
Here $\theta_{\rm max} = 85\degree$, i.e. $5\degree$ below the horizon.

\begin{figure}
\includegraphics[width=5cm, angle=270]{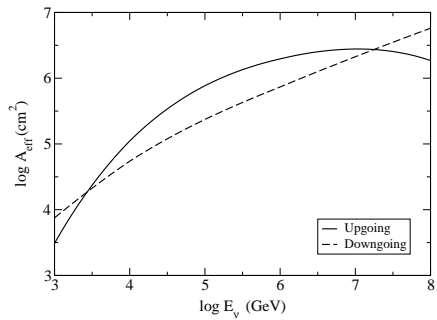} 
\caption{\label{fig:Aeff}  Estimated neutrino effective area for the detection of
upgoing and downgoing muon neutrinos
with IceCube.
For upgoing neutrinos, we have chosen $E_\mu^{\rm min} = 10^2$ GeV. For downgoing neutrinos,
we adopt the geometrical muon effective area  $A_{\mu, \rm eff} =10^{10}$ cm$^2$.}
\end{figure}

In figure \ref{fig:Aeff} we show our estimates of the IceCube effective area 
for downgoing and upgoing neutrinos,
given in eqs. 
\eqref{eq:Aeffnu:dn} and \eqref{eq:Aeffnu:up}, respectively.
The initial increase of the effective area for upgoing neutrinos  with energy is due to
the increased muon path length, whereas the subsequent decrease is due to the fact
that the Earth becomes opaque to neutrinos. These effects play no role
for downgoing neutrinos;
here the energy dependence of the
effective area follows the energy dependence of the neutrino interaction cross section.
We stress that the effective area for downgoing neutrinos corresponds
to a muon effective area  $A_{\mu, \rm eff} =10^{10}$ cm$^2$.
Given the difficulty in detecting downgoing neutrinos, this is very optimistic and
may be considered as an upper limit.

We have verified that our estimates on the neutrino effective area for
the upgoing diffuse flux agree with the estimate presented in Ref.
\cite{Desiati:2006qc} within a factor two (for energies below $10^7$ GeV)
to three (at $10^8$ GeV). Furthermore, the corresponding event rates for a fiducial 
$\phi_\nu \propto E^{-2}$ source spectrum are within a factor 2 of results
presented in Ref. \cite{Ahrens:2003ix}.

\bibliographystyle{apsrev}
\bibliography{refs}

\end{document}